\documentclass[12pt]{article}
\usepackage{epsfig}

\def\lsim{\raise0.3ex\hbox{$\;<$\kern-0.75em\raise-1.1ex\hbox{$\sim\;$}}}
\def\gsim{\raise0.3ex\hbox{$\;>$\kern-0.75em\raise-1.1ex\hbox{$\sim\;$}}}

\def\bmat{\left(\begin{array}}
\def\emat{\end{array}\right)}
\def    \be            {\begin{equation}}
\def    \ee            {\end{equation}}
\def    \bea           {\begin{eqnarray}}
\def    \eea           {\end{eqnarray}}
\def    \nn            {\nonumber}

\newsavebox{\sboxpubnumber}
\newsavebox{\sboxpubdate}
\newcommand{\pubdate}[1]{\begin{lrbox}{\sboxpubdate}{#1}\end{lrbox}}
\newcommand{\pubnumber}[1]{\begin{lrbox}{\sboxpubnumber}{\begin{tabular}{l} #1 \\
             \usebox{\sboxpubdate}
             \end{tabular}}
                           \end{lrbox}
                           \pubblock}
\newcommand{\Title}[1]{\begin{center} {\Large #1 } \end{center}}
\newcommand{\Author}[1]{\begin{center}{ \sc #1} \end{center}}
\newcommand{\Address}[1]{\begin{center}{ \it #1} \end{center}}
\newcommand{\andauth}{\begin{center}{and} \end{center}}
\newcommand{\pubblock}{\rightline{
         \usebox{\sboxpubnumber}}}
\newenvironment{Abstract}{\begin{quotation}  }{\end{quotation}}
\newenvironment{Presented}{\begin{quotation} \begin{center}
             PRESENTED AT\end{center}\bigskip
      \begin{center}\begin{large}}{\end{large}\end{center}
      \end{quotation}}
\newcommand{\Acknowledgements}{\bigskip  \bigskip \begin{center} \begin{large}
             \bf ACKNOWLEDGEMENTS \end{large}\end{center}}

\begin{document}

\begin{titlepage}
\pubdate{\today}                    
\pubnumber{CERN-TH/2001-345 \\ FTUAM 01/22 
\\ IFT-UAM/CSIC-01-38 \\ IPPP/01/57\\ DCPT/01/112} 
\vfill
\Title{Large dark matter cross sections from supergravity and superstrings}
\vfill
\Author{D. G. Cerde\~no}
%
\Address{Departamento de F\'{\i}sica
Te\'orica C-XI, Universidad Aut\'onoma de Madrid,\\
         Cantoblanco, 28049 Madrid, Spain.}
\vfill
\vfill
\Author{S. Khalil}
\Address{IPPP, Physics Department, Durham University, 
         DH1 3LE, Durham, U.K.}
\Address{Ain Shams University, Faculty of Science, Cairo, 11566, Egypt.}
\vfill
\andauth
\vfill
\Author{C. Mu\~noz}
\Address{Theory Division, CERN, CH-1211 GENEVA 23, Switzerland.}
\Address{
Departamento de F\'{\i}sica
Te\'orica C-XI and Instituto de F\'{\i}sica
Te\'orica C-XVI, Universidad Aut\'onoma de Madrid, 
Cantoblanco, 28049 Madrid, Spain.}
\vfill
\begin{Abstract}
We study the direct detection of supersymmetric dark
matter in the light of recent experimental results. In particular, 
we show that regions in the parameter space of
several scenarios with a 
neutralino-nucleon cross section of the order of $10^{-6}$ pb,
i.e., where current dark matter detectors are sensitive, can be
obtained. 
These are supergravity scenarios 
with intermediate unification 
scale, and superstring scenarios with D-branes.
\end{Abstract}
\vfill
\begin{Presented}
    COSMO-01 \\
    Rovaniemi, Finland, \\
    August 29 -- September 4, 2001
\end{Presented}
\vfill
\end{titlepage}
\def\thefootnote{\fnsymbol{footnote}}
\setcounter{footnote}{0}

\section{\large Introduction}

One of the most interesting
candidates for dark matter is a  
long-lived or stable weakly-interacting massive 
particle (WIMP). WIMPs can remain from the earliest moments of the Universe in 
sufficient number to account for a significant fraction of relic density. 
These particles would form not only a background density in the Universe, but
also would cluster gravitationally with ordinary stars in the galactic halos.

This raises the hope of detecting relic 
WIMPs directly, by observing their elastic scattering on  
target nuclei through nuclear recoils.
Since WIMPs interact with ordinary matter with very roughly weak
strength, and assuming that their masses are of the order of  
weak scale (i.e., between 10 GeV and a few TeV),
it is natural to expect a WIMP-nucleus cross section of the same order
as that of a weak process, which is around 1 pb.
This would imply a WIMP-nucleon cross section around   
$10^{-8}$ pb, too low to be detected by current dark matter
experiments,
which are sensitive to a cross section around $10^{-6}$ pb.
Surprisingly, 
the DAMA collaboration reported recently \cite{experimento1}
data favouring the existence of a 
WIMP signal in their search for annual modulation.
When 
uncertainties as e.g. the WIMP velocity or possible bulk halo rotation,
are included, it was claimed that the preferred range of parameters
is  (at 4$\sigma$ C.L.) $10^{-6}$ pb $\lsim \sigma\lsim 10^{-5}$ pb 
for a WIMP mass 30 GeV $\lsim m \lsim 200$ GeV.
However,
unlike this spectacular result, the CDMS collaboration 
claims to have excluded \cite{experimento2} regions of the DAMA 
parameter space.

Given these intriguing experimental results, it is then
crucial to re-analyze the compatibility of WIMPs as dark matter
candidates, with the sensitivity of current dark matter detectors.
To carry this analysis out we have to assume a particular candidate
for WIMP.
The leading candidate in this class is the lightest 
neutralino \cite{contemporary}, a particle 
predicted by the supersymmetric (SUSY) extension of the standard model.

In particular, in 
the 
minimal supersymmetric standard model (MSSM)
there are four neutralinos, $\tilde{\chi}^0_i~(i=1,2,3,4)$, since they
are the physical 
superpositions of the fermionic partners of the neutral electroweak 
gauge bosons, 
called bino ($\tilde{B}^0$) and wino ($\tilde{W}_3^0$), and of the
fermionic partners of the  
neutral Higgs bosons, called Higgsinos ($\tilde{H}^0_u$, 
$\tilde{H}_d^0$). 
Therefore the lightest neutralino, $\tilde{\chi}^0_1$, will be the 
dark matter candidate.
We parameterize the gaugino and Higgsino content
of the lightest neutralino according to 
\begin{equation}
\tilde{\chi}^0_1 = N_{11} \tilde{B}^0 +N_{12} \tilde{W}_3^0 +
N_{13} \tilde{H}^0_d + N_{14} \tilde{H}^0_u\ .
\label{lneu}
\end{equation}
It is commonly defined that $\tilde{\chi}^0_1$ is mostly gaugino-like 
if $P\equiv \vert N_{11}\vert^2 + \vert N_{12} \vert^2 > 0.9$, Higgsino-like
if $P<0.1$, and mixed otherwise.

The cross section for the elastic scattering of relic neutralinos on 
protons and neutrons has been examined exhaustively in the 
literature \cite{kami}.   
This is for example the case in
the framework of minimal supergravity (mSUGRA).
Let us recall that in this framework one makes several 
assumptions. In particular, 
the scalar mass parameters, the gaugino mass parameters, and the 
trilinear couplings, which are generated once SUSY is broken through
gravitational interactions, are universal at the 
grand unification 
scale, $M_{GUT} \approx 2\times 10^{16}$ GeV.
They are denoted by $m_0$, $M_{1/2}$, and $A_0$ respectively. 
Likewise, 
radiative electroweak symmetry breaking is imposed, i.e., 
the Higgsino mass parameter $\mu$ is 
determined by the minimization of the Higgs effective 
potential. This implies 
\begin{equation}
\mu^2 = \frac{m_{H_d}^2 - m_{H_u}^2 \tan^2 \beta}{\tan^2 \beta -1 } - 
\frac{1}{2} M_Z^2\ ,
\label{electroweak}
\end{equation} 
where 
$\tan\beta= \langle H_u^0\rangle/\langle H_d^0\rangle$ 
is the ratio of Higgs vacuum expectation values. 
With these assumptions, the mSUGRA framework  allows four free 
parameters: $m_0$, $M_{1/2}$, $A_0$, and $\tan \beta$. In addition, the
sign of $\mu$ 
remains also undetermined.  


It was observed (for a recent re-evaluation see ref.\cite{Ellis})
that for low and moderate values of
$\tan \beta$ the lightest neutralino is 
mainly bino, and therefore the 
predicted scalar neutralino-proton cross sections
are well below the accessible experimental regions.
In particular,
$\sigma_{\tilde{\chi}_1^0-p} \lsim 10^{-7}$ pb, and
therefore
we would have to wait in principle for projected detectors,
as e.g. GENIUS \cite{GENIUS}, 
to be able to
test the neutralino as a dark-matter candidate.

Recently, several proposals have been made in order to modify this
result, enhancing the neutralino-nucleon cross section.
This is the case of scenarios with large 
$\tan \beta$ \cite{Bottino}-\cite{Mario}, with
non-universal soft SUSY-breaking terms \cite{Bottino,Arnowitt,Nath2}, 
with multi-TeV masses
for scalar superpartners known as `focus point' 
supersymmetry \cite{focus}, 
with intermediate unification scale \cite{muas},
and
finally superstring scenarios with D-branes 
\cite{khalil,Nath2,Arnowitt2,bailin,nosotros}. 
In these proceedings we will concentrate only
on the last two.
A recent review where all the above mentioned scenarios are critically
reappraised can be found in ref.\cite{darkcairo}.

\section{\large Scenarios with intermediate unification scale}

The analyses of the neutralino-nucleon cross section in mSUGRA, 
mentioned above, were performed assuming the
unification scale
$M_{GUT} \approx 10^{16}$ GeV, as is usually done in the SUSY literature.
However, it was recently 
pointed out \cite{muas} 
that this
cross section 
$\sigma_{\tilde{\chi}_1^0-p}$
is very sensitive to the variation of the initial scale 
for the running of the soft terms.
In particular, intermediate unification scales were considered.
For instance, by taking $M_I=10^{10-12}$ GeV rather than 
$M_{GUT}$, regions in the parameter space of mSUGRA have been 
found 
where  $\sigma_{\tilde{\chi}_1^0-p}$
is in the expected range of sensitivity of present detectors,
and this
even for moderate values of $\tan\beta$ ($\tan\beta\gsim 3$). 
This analysis 
was performed in the universal scenario for the soft terms. 
In contrast, in the usual case with initial scale at $M_{GUT}$,
this large cross section is achieved only for $\tan\beta \gsim 20$.

\begin{figure}[t]
\begin{center}
\epsfig{file= 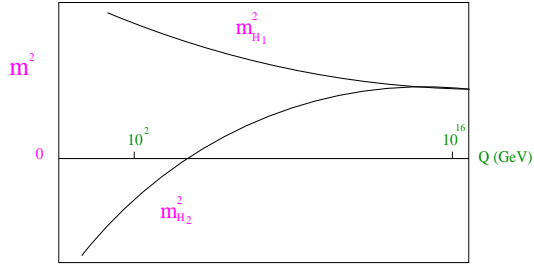,width=7cm,height=3.5cm,angle=0}
\end{center}
\caption{
Running of the soft Higgs masses-squared with energy $Q$.}
\label{run}
\end{figure}

The fact that smaller initial scales imply a larger neutralino-proton
cross section can be understood from the variation in the value of 
$\mu$ with $M_I$.
One observes that, for $\tan\beta$ fixed, the smaller the initial
scale for the running is, the smaller the numerator in the
first piece of eq.(\ref{electroweak}) becomes. 
This can be understood qualitatively from Fig.~\ref{run}, where
the well known evolution of $m_{H_d}^2$ and $m_{H_u}^2$ with the
scale is schematically shown. 
Clearly, the smaller the initial scale is, the shorter the
running becomes. As a consequence, 
also the less important the positive(negative) contribution 
$m_{H_d}^2$($m_{H_u}^2$) to $\mu$ in eq.(\ref{electroweak}) becomes.
Thus 
$|\mu|$ decreases. 

\begin{figure}[ht]
\begin{center}
\epsfig{file= 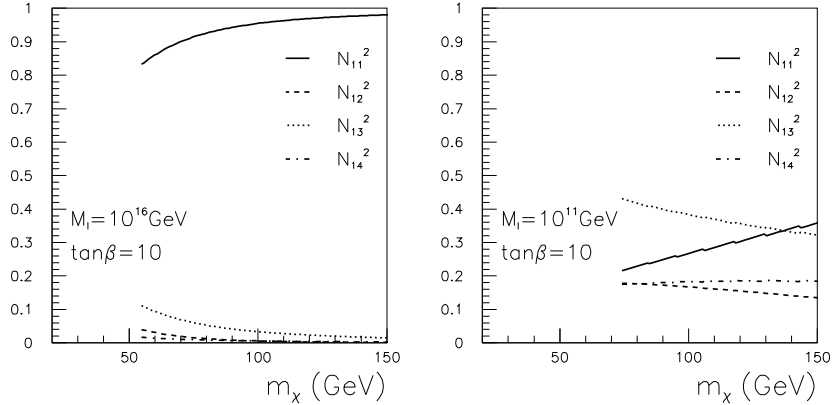, 
height=7cm,angle=0}
\end{center}
\vspace{-0.5cm}
\caption{Gaugino-Higgsino components-squared
of the lightest neutralino as a function of its mass for
the unification scale, $M_I=10^{16}$ GeV, and for the intermediate scale,
$M_I=10^{11}$ GeV.
}
\label{N_1i}
\end{figure}

\begin{figure}[t]
\begin{center}
\epsfig{file= 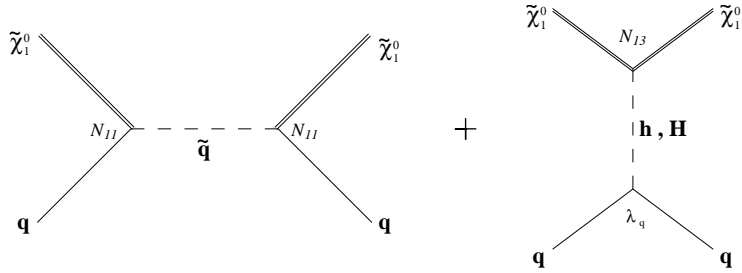,height=3.5cm,angle=0}
\end{center}
\caption{Feynman diagrams contributing to neutralino-nucleon cross section.}
\label{Feynman}
\end{figure}

Now, since 
${\cal L} \sim \mu \tilde{H}^0_u \tilde{H}^0_d$ + h.c.,
the Higgsino
components
of the lightest neutralino, $N_{13}$ and $N_{14}$ in eq.(\ref{lneu}),
increase.
In Fig.~\ref{N_1i}, for $\tan\beta=10$ and $m_0=150$ GeV, 
we exhibit the gaugino-Higgsino components-squared $N_{1i}^2$ 
of the lightest neutralino as a function of its mass $m_{\tilde{\chi}_1^0}$
for 
two different values of 
the initial scale, 
$M_I=10^{16}$ GeV  $\approx M_{GUT}$ and 
$M_I=10^{11}$ GeV. 
Clearly, the smaller the scale is, the larger 
the Higgsino components become. 
%
%
For
$M_I=10^{11}$ GeV, e.g. the Higgsino contribution $N_{13}$  
becomes important 
and even dominant for 
$m_{\tilde{\chi}_1^0}\lsim 140$ GeV.
Then, the scattering channels through Higgs exchange shown in 
Fig.~\ref{Feynman} are important,
and therefore
the cross section may be large.

%
%

\newpage

\begin{figure}[ht]
\begin{center}
\epsfig{file= 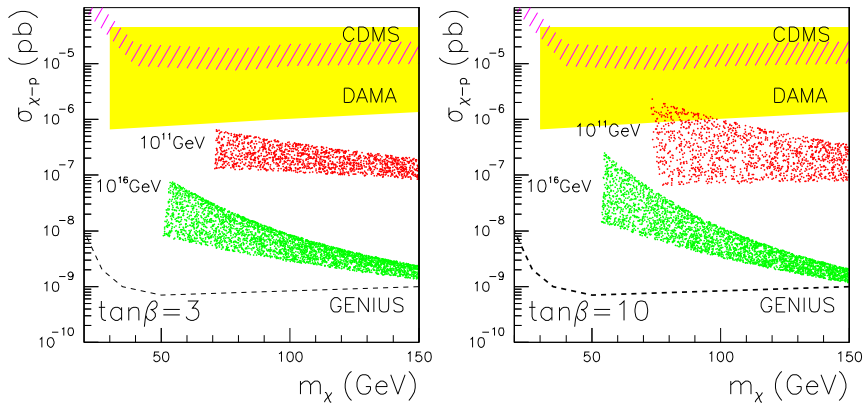, 
height=7cm}
\end{center}
\vspace{-0.5cm}
\caption{Scatter plot of the neutralino-proton cross section
as a function of the neutralino mass for two values of the initial scale,
$M_I=10^{16}$ GeV and $10^{11}$ GeV, and for $\tan\beta=3$ and 10.
DAMA and CDMS current 
experimental limits and projected GENIUS limits are shown.
}
\label{sigmaM_I}
\end{figure}

This is shown in 
Fig.~\ref{sigmaM_I}, where
the cross section as a function
of the lightest neutralino mass $m_{\tilde{\chi}_1^0}$ is plotted.
In particular we are comparing the result for the scale $M_I=M_{GUT}$
with the result for 
the intermediate scale $M_I=10^{11}$ GeV.
For instance, when $m_{\tilde{\chi}_1^0}=100$ GeV, 
$\sigma_{\tilde\chi_1^0-p}$ 
for $M_I=10^{11}$ GeV is two orders of magnitude larger
than for $M_{GUT}$.
In particular, for $\tan\beta=3$, one finds 
$\sigma_{\tilde\chi_1^0-p} \lsim 10^{-7}$ pb 
if the initial scale is $M_I=10^{16}$ GeV.
However $\sigma_{\tilde\chi_1^0-p} \lsim 10^{-6}$ GeV is possible if 
$M_I$ decreases. 


It is also worth noticing
that, for any fixed value of $M_I$, the larger
$\tan\beta$ is, the larger
the Higgsino contributions become,
and
therefore the cross section increases.
For $\tan\beta=10$ we see in Fig.~\ref{sigmaM_I}
that the range 70 GeV $\lsim m_{\tilde{\chi}_1^0}\lsim$ 100 GeV is now
consistent 
with DAMA limits. 

%


Let us remark that these figures have been obtained \cite{muas}
taking $30\lsim m_0\lsim 550$ GeV and $A_0=M_{1/2}$.
In any case, the cross section is 
not very sensitive to the specific values
of $A_0$. In particular it was checked that 
this is so for 
$\mid A_0/M_{1/2}\mid \lsim 1$.

Let us finally recall that
non universality of the soft terms in 
addition to intermediate scales
may introduce more flexibility in the computation.
In particular, decreasing $|\mu|$
in order to obtain regions in the parameter space giving rise to  
cross sections compatible with the sensitivity
of current detectors, may be easier.

\section{\large Superstring scenarios}

In the above section the analyses were performed assuming intermediate
unification scales.
In fact, this situation can be inspired by superstring theories, since
it was recently 
realized that 
the string scale may be anywhere between the weak and the Plank 
scale \cite{Lykken}-\cite{weakandstronghete}. 
For example, embedding the standard model inside D3-branes in type I
strings, the string scale is given by:
\bea
M_I^4= \frac{\alpha M_{Planck}}{\sqrt 2} M_c^3\ ,
\label{gravitino2}
\eea
where $\alpha$ is the gauge coupling and $M_c$ is the compactification scale. 
Thus one gets $M_I\approx 10^{10-12}$ GeV with $M_c\approx 10^{8-10}$ GeV.

Then, to use the value of the initial scale,
say $M_I$, as a free parameter for the running of the soft terms
is particularly interesting.
In addition, 
there are several arguments in favour of SUSY scenarios
with scales 
$M_I\approx 10^{10-14}$ GeV.
These scales were suggested \cite{stronghete,typeIinter}
to explain many experimental observations as
neutrino masses or the scale for axion physics. 
With the string scale of the order of
$10^{10-12}$ GeV
one is also able
to attack the hierarchy problem of unified theories 
without invoking any hierarchically suppressed non-perturbative
effect \cite{stronghete,typeIinter}. 
In supergravity models supersymmetry can be spontaneously broken in a 
hidden sector of the theory and the gravitino mass,
which sets the overall scale of the soft terms, is given by
$m_{3/2}\approx \frac{F}{M_{Planck}}$,
where $F$ is the auxiliary field whose vacuum expectation value
breaks supersymmetry. 
Since in supergravity one would expect $F\approx M_{Planck}^2$, one  
obtains
$m_{3/2}\approx M_{Planck}$ and therefore 
the hierarchy problem solved in principle by supersymmetry
would be re-introduced.
However, if the scale
of the fundamental theory is $M_I\approx 10^{10-12}$ GeV instead of
$M_{Planck}$,
then $F\approx M_I^2$
and one gets $m_{3/2}\approx M_W$ in a natural way.

There are other arguments in favour 
of scenarios with initial scales $M_I$ smaller than $M_{GUT}$.
For example,
charge and color breaking constraints,
which are very strong with the usual scale $M_{GUT}$ \cite{servidor},
become less important \cite{Allanach}.
These scales might also
explain the observed ultra-high energy ($\approx 10^{20}$ eV) cosmic rays
as products of long-lived massive string mode decays.
\cite{stronghete,typeIinter} (see ref.\cite{cosmic} for more
details about this possibility). Besides,
several models of chaotic inflation favour also these scales \cite{caos}.

\begin{figure}[ht]
\begin{center}
\begin{tabular}{c}
\epsfig{file= 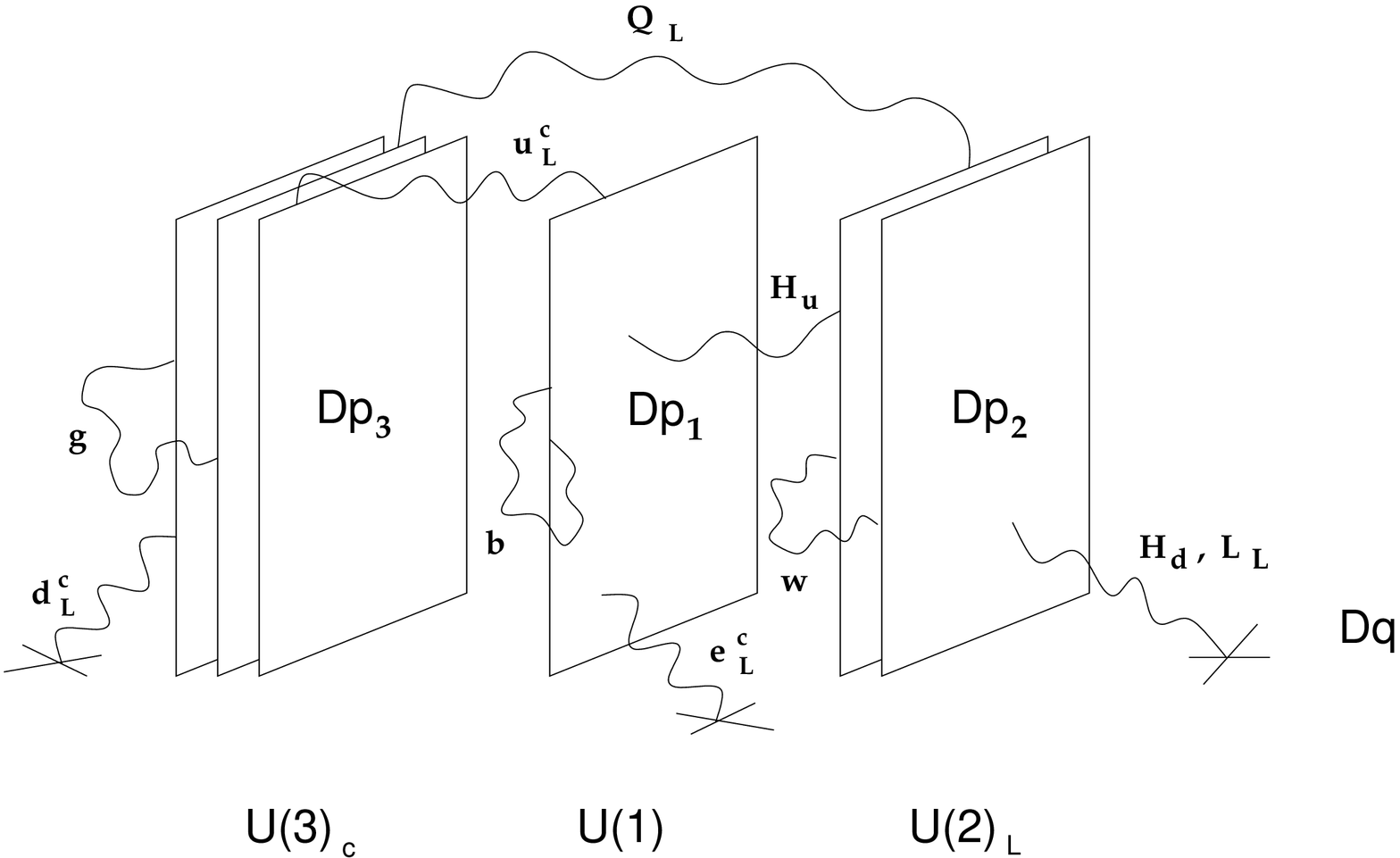, width=9cm}\\
\end{tabular}
\end{center} 
\caption{A generic D-brane scenario giving rise to the gauge bosons
and matter of the standard model. It contains three D$p_3$-branes,
two D$p_2$-branes and one D$p_1$-brane, where $p_N$ may be
either 9 and $5_i$ or 3 and $7_i$. The presence of extra D-branes,
say D$q$-branes, is also
necessary. For each set the D$p_N$-branes
are in fact on the top of each other.} 
\label{branas}
\end{figure}

D-brane constructions are explicit scenarios where the two situations 
mentioned above, namely,
non-universality and intermediate scales, may occur.
Let us then analyse this possibility.

\subsection{\large D-branes}

The first attempts to study dark matter within these constructions
were carried out in scenarios with the unification scale  
$M_{GUT} \approx 10^{16}$ GeV
as the initial scale \cite{khalil,Nath2,Arnowitt2}
and dilaton-dominated SUSY-breaking scenarios 
with an intermediate scale as the initial scale \cite{bailin}.
However,
the important issue of the D-brane origin of the $U(1)_Y$ gauge group
as a combination of other $U(1)$'s 
and its influence on the matter distribution in these scenarios
was not included in the above analyses.
When this is taken into account, interesting results are 
obtained \cite{nosotros}. In particular, 
scenarios with the gauge group and particle content of the
SUSY standard model lead naturally to intermediate values for the
string
scale, in order to reproduce the value of gauge couplings
deduced from experiments. In addition, the soft terms 
turn out to be generically non universal.
Due to these results, 
large cross sections
in the small $\tan\beta$ regime
can be obtained.

Let us consider for example a type I string scenario \cite{nosotros}
where the gauge group 
$U(3)\times U(2)\times U(1)$, giving rise to
$SU(3)\times SU(2)\times U(1)^3$, arises
from three different types of D-branes, 
as shown schematically in Fig.~\ref{branas}.
For the sake of visualization each set is
depicted at parallel locations, but in fact they are intersecting each
other.
Other examples with the standard model gauge group embedded in D-branes in 
a different way can be found in ref.\cite{nosotros}.
Here
$U(1)_Y$ is a linear combination
of the three $U(1)$ gauge groups arising from $U(3)$, $U(2)$ and
$U(1)$ within the three different D-branes. 
\begin{figure}[t]
\begin{center}
\begin{tabular}{c}
\epsfig{file= 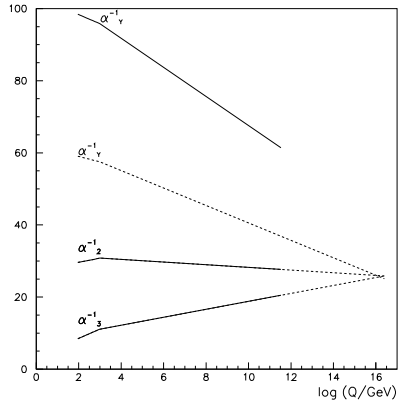, width=6cm}\\
\end{tabular}
\end{center} 
\caption{Running of the gauge couplings of the MSSM with energy $Q$ 
embedding the gauge groups within different sets of D$p$-branes (solid lines).
Due to the D-brane origin of the $U(1)$ gauge groups, relation
(\ref{couplings}) must be fulfilled. 
For comparison
the running of the MSSM couplings with
the usual normalization factor for the hypercharge, $3/5$,
is also shown with dashed lines.} 
\label{corriendo}
\end{figure}
This implies
\bea
Y=-\frac{1}{3} Q_3 - \frac{1}{2} Q_2+ Q_1\ ,
\label{hypercharge}
\eea
and therefore, 
\begin{equation}
\frac{1}{\alpha_Y(M_I)} =     
\frac{2}{\alpha_1(M_I)} + \frac{1}{\alpha_2(M_I)} 
+ \frac{2}{3 \alpha_3(M_I)}\ ,
\label{couplings}
\end{equation}
where $\alpha_k$ correspond to the gauge couplings of the $U(k)$
branes.
Let us remark that since the D-branes are of different types,
the gauge couplings are
non-universal.
Now, using the RGEs for gauge couplings one obtains
\bea
\ln\frac{M_I}{M_s}=
\frac{2\pi\left(\frac{1}{\alpha_Y (M_Z)}-\frac{2}{\alpha_1 (M_I)}-
\frac{1}{\alpha_2 (M_Z)}-
\frac{2}{3\alpha_3 (M_Z)}\right)
+\left(b_Y^{ns}-b_2^{ns}-\frac{2}{3} b_3^{ns}\right)
\ln\frac{M_s}{M_Z}
}
{\left(\frac{2}{3} b_3^s + b_2^s -b_Y^s\right)}
\ ,
\label{MI}
\eea
where $b_j^s$ ($b_j^{ns}$) with $j=2,3,Y$ are the coefficients of 
the supersymmetric (non-supersym\-metric) $\beta$-functions, and 
the scale $M_{s}$ corresponds to the supersymmetric threshold, 
200 GeV $\lsim M_s\lsim$ 1000 GeV. 
For example, choosing the value of the coupling associated 
to the D$p_1$-brane in the range $0.07\lsim\alpha_1 (M_I)\lsim 0.1$,
and the experimental values for $\alpha_{3,2,Y}$,
one obtains
$M_I\approx 10^{10-12}$ GeV. This scenario is shown in Fig.~\ref{corriendo}
for $\alpha_1 (M_I)=0.1$ and $M_s=1$ TeV.
%
Let us remark that 
the extra $U(1)$'s are anomalous and therefore the associated gauge
bosons
have masses of the order of $M_I$.

The analysis of the soft terms has been done under
the assumption that only the
dilaton ($S$) and moduli ($T_i$) fields contribute to SUSY
breaking and it has been found  that these soft terms  are generically
non-universal.
Using the standard parameterization \cite{dilaton} 
\begin{eqnarray}
&& F^S= \sqrt{3} (S+S^*) m_{3/2} \sin \theta\;, \nonumber \\
&& F^i= \sqrt{3} (T_i+T^*_i) m_{3/2} \cos \theta\; \Theta_i\;,
\end{eqnarray}
where
$i=1,2,3$ labels the three complex compact dimensions, and
the angle $\theta$ and the $\Theta_i$ with $\sum_{i} |\Theta_i|^2=1$,
just parameterize the direction of the goldstino in the $S$, $T_i$ field
space, one is able to obtain the following soft terms \cite{nosotros}.
The gaugino masses associated to the three gauge groups of the
standard model are given by
%
\bea
M_3 & = & \sqrt{3} m_{3/2} \sin \theta \ , \nn\\
M_{2} & = & \sqrt{3}  m_{3/2}\ \Theta_1 \cos \theta  \ , \nn\\
M_{Y} & = &  \sqrt{3}  m_{3/2}\ \alpha_Y (M_I)
\left(\frac{2\ \Theta_3 \cos \theta}{\alpha_1 (M_I)}
+\frac{\Theta_1 \cos \theta}{\alpha_2 (M_I)}
+\frac{2\ \sin \theta}{3 \alpha_3 (M_I)}
\right)\ .
\label{gaugino1}
\eea
The soft scalar masses of the three families are given by
\begin{eqnarray}
m^2_{Q_L} & = & m_{3/2}^2\left[1 -
\frac{3}{2}  \left(1 - \Theta_{1}^2 \right)
\cos^2 \theta \right] \ , \nn \\
m^2_{d_R} & = & m_{3/2}^2\left[1 -
\frac{3}{2}  \left(1 - \Theta_{2}^2 \right)   
\cos^2 \theta \right] \ , \nn \\
m^2_{u_R} & = & m_{3/2}^2\left[1 -
\frac{3}{2}  \left(1 - \Theta_{3}^2 \right)
\cos^2 \theta \right] \ , \nn \\
m^2_{e_R} & = & m_{3/2}^2\left[1- \frac{3}{2}
\left(\sin^2\theta + \Theta_{1}^2 \cos^2\theta  \right)\right] \ , \nn \\
m^2_{L_L} & = & m_{3/2}^2\left[1- \frac{3}{2}
\left(\sin^2\theta + \Theta_{3}^2 \cos^2\theta  \right)\right] \ , \nn \\
m^2_{H_u} & = & m_{3/2}^2\left[1- \frac{3}{2}
\left(\sin^2\theta + \Theta_{2}^2 \cos^2\theta  \right)\right] \ , \nn \\
m^2_{H_d} & = & m^2_{L_L} \;,     
\label{scalars1}
\end{eqnarray}
where e.g. $u_R$ denotes the three family squarks $\tilde{u}_R$, 
$\tilde{c}_R$, $\tilde{t}_R$.
Finally the trilinear parameters of the three families are
\begin{eqnarray}
A_{u} & = &  \frac{\sqrt 3}{2}m_{3/2}
   \left[\left(\Theta_{2} - \Theta_1
 - \Theta _{3}  \right) \cos\theta
- \sin\theta \; \right] \ ,
\nn \\
A_{d} & = &  \frac{\sqrt 3}{2}m_{3/2}
   \left[\left(\Theta_{3} - \Theta_1
  - \Theta _{2}  \right) \cos\theta
- \sin\theta \; \right] \ ,
\nn \\
A_{e} & = &  0\; .
\label{trilin11}
\end{eqnarray}
\begin{figure}[t]
\begin{center}
\epsfig{file= 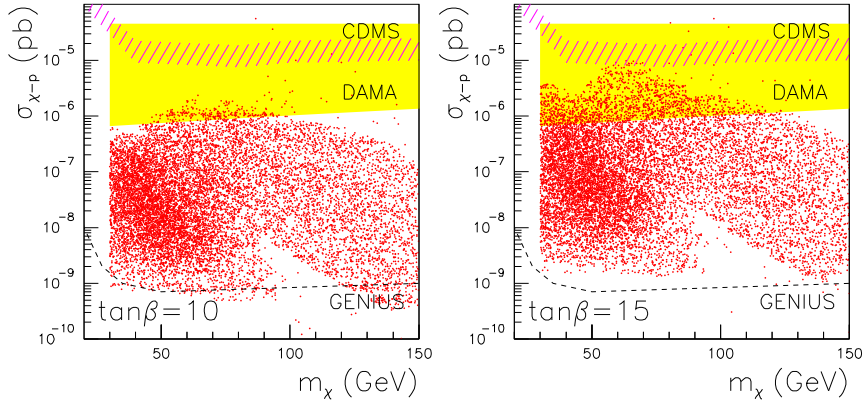, 
height=7cm
}
\end{center}
\vspace{-1cm}
\caption{The same as in Fig.~\ref{sigmaM_I} but for 
the D-brane scenario with 
the string scale $M_I=10^{12}$ GeV
discussed
in the text, and for $\tan\beta= 10$ and 15.}
\label{dbrane}
\end{figure}

Although these formulas for
the soft terms imply that one has in principle
five free parameters, $m_{3/2}$, $\theta$ and  $\Theta_i$ with $i=1,2,3$,
due to relation $\sum_i |\Theta_i|^2=1$ only four of them are
independent.
In the analysis the parameters $\theta$ and $\Theta_i$
are varied in the whole allowed range, $0\leq \theta\leq 2\pi$,
$-1\leq\Theta_i\leq 1$. For 
the gravitino mass, 
$m_{3/2}\leq 300$ GeV is taken.
Concerning Yukawa couplings, their values are fixed imposing 
the correct fermion mass spectrum at low energies, i.e.,
one is assuming that Yukawa structures of D-brane scenarios
give rise to those values.

Fig.~\ref{dbrane} displays a scatter plot of 
$\sigma_{\tilde\chi_1^0-p}$ as a function of the
neutralino mass $m_{\tilde\chi_1^0}$ for a scanning of the parameter space
discussed above. 
Two different values of $\tan\beta$, 10 and 15, are shown.
It is worth noticing that
for $\tan\beta =10$ there are regions of the parameter
space consistent with DAMA limits.
In fact, one can check that $\tan\beta > 5$ is enough to
obtain compatibility with DAMA.
Since the larger $\tan\beta$ is, the
larger the cross section becomes, for $\tan\beta =15$ these regions
increase.




\section{\large Relic neutralino density versus cross section}

As discussed in the Introduction, current dark matter detectors
are sensitive to a neutralino-proton cross section around $10^{-6}$ pb.
This value is obtained taking into account, basically, that the
density of dark matter in our Galaxy, which follows from the 
observed rotation curves, is 
$\rho_{DM}\approx 0.3$ GeV/cm$^3$.
Thus in this work we were mainly interested in reviewing the
possibility of obtaining such large cross sections in the
context of mSUGRA and superstring
scenarios. In order to compute the cross section
only simple field theory techniques are needed,
no cosmological assumptions about the
early Universe need to be used.

On the other hand, such
cosmological assumptions indeed must be taken into account 
when computing the amount of relic neutralino density
arising from the above scenarios.
Generically, through thermal production of neutralinos, one obtains \cite{kami}
%
\begin{equation}
\Omega_{\tilde{\chi}_1^0} h^2 \simeq \frac{C}
{<\sigma_{\tilde{\chi}_1^0}^{ann}. v>}
\ ,
\label{assumption}
\end{equation}
where $\sigma_{\tilde{\chi}_1^0}^{ann}$ is the  
cross section for annihilation of a pair of neutralinos
into standard model particles, $v$ is the 
relative velocity between the two neutralinos, 
and $<..>$ denotes thermal averaging.
The constant $C$ involves factors of Newton's constant, the temperature of the 
cosmic background radiation, etc. 
Then one may compare this result with dark matter observations
in the Universe.
Let us then discuss briefly the effect of relic neutralino 
density bounds on cross sections.

The most robust evidence for the existence of dark matter comes from 
relatively small scales, in particular,
from the flat rotation curves of spiral Galaxies. 
On the opposite side, observations at large scales,
have also provided 
estimates 
of $\Omega_{DM}$.
Taking 
both kind of observations
one is able to obtain a favoured 
range \cite{contemporary}
$0.1\lsim \Omega_{DM} h^2 \lsim 0.3$,
where $h$ is the reduced Hubble constant. 
It is worth noticing, however, that more conservative
lower limits have also been quoted in the literature (a
brief discussion can be found in ref.\cite{darkcairo} and references therein).

As is well known, for $\sigma_{\tilde{\chi}_1^0}^{ann}$ 
of the order 
of a weak-process cross section, $\Omega_{\tilde{\chi}_1^0}$ obtained
from eq.(\ref{assumption}) is 
within the favoured range discussed
above \cite{kami}. 
This is precisely the generic case when the lightest neutralino
is mainly bino. Then, the neutralino-nucleus cross section is
of the order of 1 pb,
i.e. $\sigma_{\tilde{\chi}_1^0-p}\approx 10^{-8}$ pb, 
and therefore it is natural to obtain
that neutralinos annihilate with very roughly
the weak interaction strength. 
In fact, for these cross sections, there is always
a set of parameters which yield $0.1<\Omega_{\tilde\chi_1^0}h^2<0.3$.
This analysis, including a complete treatment of coannihilations
was carried out in refs.\cite{Ellis,Ellisco}.

On the other hand, in these proceedings we were interested in larger
neutralino-nucleon cross sections in order to be in the range
of sensitivity of current dark matter detectors.
It is then expected that such high neutralino-proton cross sections
$\sigma_{\tilde\chi_1^0-p}\approx 10^{-6}$ pb,
as those presented in Sections~2 and 3, will correspond to relatively low
relic neutralino densities.
This is in fact the general situation 
\cite{muas,bailin,nosotros}.

Let us remark, however,
that thermal production of neutralinos is not the only possibility,
moduli decays can also produce neutralinos.
Since the decay width of the moduli is
$\Gamma_{\phi}\sim \frac{m_{\phi}^3}{M^2}$, 
scenarios with intermediate scales 
might give rise to cosmological
results different from the usual ones 
summarized in eq.(\ref{assumption}).
Note that if $M=M_I$ the usual moduli problem 
may be avoided since a reheating temperature $T_{RH}$ small but 
larger than 1 MeV
can be obtained.
This is e.g. the case of the twisted moduli fields 
$M_{\alpha}$ in type I strings.
Recall that
$n/s
\propto 1/T$
and since we can have a situation with 
$T_{RH}<<T_f\approx m_{\chi}/20\sim \cal{O}$(1 GeV), 
the relic neutralino density might  be larger than in the
usual case of thermal production. As a consequence,
$\Omega_{\tilde\chi_1^0}h^2> 0.1$ may be obtained \cite{preparation}.

\section{\large Final comments and outlook}

In the present proceedings we have studied  the direct detection of 
supersymmetric dark matter in the light of recent experimental efforts.
In particular, DAMA collaboration using a NaI detector
has reported recently 
data favouring the existence of a WIMP signal in their search for
annual modulation. They require a large cross section of the order of
$10^{-6}$ pb. 
We have observed that there are regions in the parameter space
of mSUGRA scenarios with intermediate scales and
superstring scenarios with D-branes 
where such a value can be obtained, although it is fair to say that
smaller values
can also be obtained and even more easily.
The latter result may be important since 
CDMS collaboration using a germanium detector
has reported a null result for part of the region explored by DAMA. 
Clearly, more sensitive detectors
producing further data are needed to solve this contradiction.
Fortunately, many dark matter detectors are being projected.
This is the case e.g. of DAMA 250 kg., CDMS Soudan, GENIUS, etc.
where values of the cross section as low as 
$10^{-9}$ pb will be accesible. 

In summary, underground physics as the one discussed here in order
to detect dark matter is crucial. Even if neutralinos are discovered
at future particle accelerators such as LHC, only their direct detection
due to their presence in our galactic halo will confirm that they
are the sought-after dark matter of the Universe.

\Acknowledgements

\noindent
We thank E. Gabrielli and E. Torrente-Lujan
as co-authors of some of the works reported in
this review. D.G. Cerde\~no acknowledges the financial support
of the Comunidad de Madrid through a FPI grant.
The work of S. Khalil was supported by PPARC.
The work of C. Mu\~noz was supported in part by the Ministerio de 
Ciencia y Tecnolog\'{\i}a (Spain), and the European Union under contract 
HPRN-CT-2000-00148. 


\end{document}